\documentclass[aps,pra,showpacs,twocolumn,amsmath,amssymb,footinbib]{revtex4-1}

\usepackage{latexsym}
\usepackage{graphicx}
\usepackage{subfigure}
\usepackage{epsfig}
\usepackage{amsfonts}
\usepackage{amssymb}
\usepackage{amsmath}
\usepackage{bbm}

\usepackage{tikz}
\usetikzlibrary{arrows,positioning,matrix}
	\newcommand{\midarrow}{\tikz \draw[-triangle 90] +(.1,0) -- (0,0);}

\usepackage[colorlinks,bookmarks=false,citecolor=blue,linkcolor=red,urlcolor=blue]{hyperref}

\begin{document}

\title{Nontrivial ground-state degeneracies and\\generalized fractional excitations in the 1D lattice}

\author{Emma Wikberg}

\affiliation{Department of Physics,
Stockholm University, AlbaNova University Center, 106 91 Stockholm,
Sweden}

\date{\today}

\begin{abstract}
We study a 1D lattice Hamiltonian, relevant for a wide range of interesting physical systems like, e.g., the quantum-Hall system, cold atoms or molecules in optical lattices, and TCNQ salts. Through a tuning of the interaction parameters and a departure from a strictly convex interaction, nontrivial ground-state degeneracies and fractionally charged excitations emerge. The excitations, being a generalization of the fractional charges known from the fractional quantum-Hall effect, appear as domain walls between inequivalent ground states and carry the charge $\pm\frac{e}{mq}$, where $m$ is an integer and associated with the specified interaction, and $\nu=p/q$ is the filling fraction in the lattice. The description points at an interesting resemblance to states connected to non-Abelian statistics, which is central for the concept of topological quantum computing.
\end{abstract}

\pacs{67.85.-d, 71.10.Pm, 75.10.Pq}
\maketitle


\section{Introduction and motivation}

The electrostatic problem in 1D has a huge number of different applications, ranging from earlier studies of tetracyanoquinodimethane (TCNQ) salts \cite{hubbard} to more modern examples like optical lattices.  Within the latter area, a fast and interesting development has taken place during the last decade \cite{lewenstein,bloch}. 1D lattice systems, in which the interaction parameters are sensitive to tuning, open the thrilling possibility to realize and explore different phases of matter under controllable circumstances. Assuming the approximation that there is no tunneling between sites \cite{com1}, these systems are described by the Hamiltonian
\begin{equation}
\label{eqn:ham}
\hat{H}=\frac{V_0}{2}\sum_{i=1}^\infty \hat n_i(\hat n_i-1)+\sum_{i,j=1}^\infty V_j\hat n_i\hat n_{i+j},
\end{equation} 
where $V_0$ is the energy cost of a pair of particles occupying the same site, $V_j$ is the energy cost of a pair of particles $j$ lattice sites apart, and $\hat n_i$ counts the number of particles on site $i$. As it stands, this Hamiltonian is independent of particle statistics, apart from the fact that when the particles in question are fermions, $n_i$ is either zero or one. 

Beyond the comparably obvious fields of application already mentioned, the 1D lattice model turns out to be, somewhat more surprisingly, relevant for completely different systems. One example is the fractional quantum-Hall (FQH) system \cite{tsui}, which is two-dimensional in the laboratory but reveals its one-dimensional nature when mathematically mapped onto a cylinder or a torus, see \cite{emilPRB} and references therein. In the limit where the thickness of the torus approaches zero, the system is described by equation \eqref{eqn:ham}. Furthermore, the quantum-Hall (QH) system is mathematically equivalent to that of rotating neutral bosons in zero magnetic field \cite{susanne}, which adds yet another interesting system to the list of those being captured by $\hat H$ above.

The relevance of \eqref{eqn:ham} for a large diversity of physical systems certainly motivates a thorough analysis of this Hamiltonian. It turns out that the low-energy behavior is crucially dependent on the signs of the ``second derivatives" of the discrete set of interaction parameters $V_j$, defined as $V''_j\equiv V_{j-1}+V_{j+1}-2V_j$, $j\geq1$. While many valuable insights have been gained before concerning strictly convex interactions ($V''_j>0$ $\forall j$) in general \cite{hubbard,pokrovsky,emilPRB,emilPRL,burnell}, as well as some special cases of \emph{non-convex} dittos (meaning $V''_j\geq0$) \cite{bosonartikeln,jonas,burnell,bauer}, we here analyze the 1D lattice system for more general non-convex interactions, exploring the possible outcomes of tuning the parameters $V_j$. 

It has earlier been shown that for any convex interaction and particle filling fraction $\nu=p/q$, the ground state is the state where the particles are as far separated as possible. The state has a (trivial) $q$-fold center-of-mass degeneracy and the minimal excitations, carrying charges $\pm\frac{e}{q}$, are formed as domain walls between translated ground states \cite{emilPRB,su}. In this paper we show that for certain choices of non-convex interactions, there is a \emph{nontrivial} $q(2^m-1)$-fold ground-state degeneracy and the lowest excitations---analogously formed as domain walls between the degenerate vacua---have fractional charges which are even smaller than the ones resulting from an ordinary convex interaction. Here, $m$ is an integer associated with the specified interaction and a generalization of the Su-Schrieffer counting argument \cite{su} shows that the charges are of size $\pm\frac{e}{mq}$. A nontrivial ground-state degeneracy is a prerequisite for the realization of non-Abelian statistics \cite{stern}, having connections to topological quantum computing \cite{nayak}, and our solution is relevant for, e.g., the non-Abelian Moore-Read state \cite{mr} known from quantum-Hall physics.

The outline of the paper is as follows. In Section \ref{sec2} we review previously known results concerning the properties of the 1D lattice system for ordinary convex interactions, as well as discuss what may be expected in general when one infers a departure from convexity and assumes different kinds of \emph{non-convex} interactions. Following that, in Section \ref{sec3} we present the results of our analysis, valid for a certain class of non-convex interactions. The nature of ground states and minimal excitations is explored in Sections \ref{sub1} and \ref{subB} respectively, followed by reflections on experimental possibilities and other implications in \ref{subexp}. Finally, a summary is enclosed in Section \ref{sec5}. For further details concerning the analysis, which employs a mapping of the 1D lattice system onto spin-1/2 chains, we refer the reader to the Appendix.


\section{Convex and non-convex interactions}\label{sec2}

\subsection{Convexity and charge fractionalization}\label{subconvex}

As mentioned in the introduction, the properties of a system described by equation \eqref{eqn:ham} depend on the signs of the interaction parameters $V''_j\equiv V_{j-1}+V_{j+1}-2V_j$, $j\geq1$. One important class of interactions, including the ordinary screened and unscreened Coulomb, and dipole-dipole interaction ($V_j=V_1/j^3$), contains those for which $V''_j>0$ and $V_j>V_{j+1}$, $\forall j$. For obvious reasons, these are denoted \emph{convex interactions}. This case has been studied extensively \cite{burnell,hubbard,emilPRL,emilPRB,pokrovsky} and the solution is well known; the ground state at particle filling fraction $\nu=p/q$, $p$ and $q$ being integers and coprime, is the state where the particles are as evenly spread out as possible on the lattice, yielding a unit cell of length $q$ and consequently a trivial, $q$-fold center-of-mass degeneracy. For example, the ground state at $\nu=2/5$ has unit cell $[01010]$ (translations $[00101]$, $[10010]$, $[01001]$, and $[10100]$) and $\nu=3/2$ has unit cell $[21]$ (translation $[12]$). These types of states will be referred to as Tao-Thouless (TT) states \cite{tt} and, in what follows, the bracket notation will be used to denote both the unit cells themselves as well as entire states, depending on the context. To be specific, for some purposes $...121212...\equiv[12]$.  


As for the excitation structure in the case of convex interactions, the lowest excitations are generically created through the formation of domain walls between degenerate TT ground states of the specific particle filling fraction \cite{emilPRB,emilPRL,burnell} and the so-called Su-Schrieffer counting argument \cite{su} shows that the domain walls carry fractional charge $\pm\frac{e}{q}$, where $e$ is the charge of the particles themselves. (The equivalent for neutral particles is fractional particle number $\pm\frac{1}{q}$.) An alternative way to do the counting \cite{eddy} is to look at the number of particles in each string of $q$ consecutive sites. In the excited state, all such strings host exactly $p$ particles, except for the string at the domain wall, which carries $p\pm 1$ (depending on how the domain wall is constructed) particles. Compared to the background, the domain wall thus carries an extra charge $\pm e/q$. 

Due to this rule, there is a restriction on which translated copies of the TT state that may be combined to create a minimal excitation. The unit cell of a TT state may be written in the form $[\tilde C10]$, where $\tilde C$ is a binary string, mirror symmetric around its center \cite{comCtilde,micke}. For $\nu=2/5$, e.g., $\tilde C=010$. Furthermore, the cell $[\tilde C01]$, which one gets by moving the rightmost particle one step to the right, is a translated copy of the unit cell one started with; for $\nu=2/5$, $[01010]\rightarrow[01001]$.  The nature of the fractional charge $\pm\frac{e}{q}$ is a domain wall between these two unit cells; the domain wall $[\tilde C10][\tilde C01]$ corresponds to a quasi-hole and $[\tilde C01][\tilde C10]$ corresponds to a quasi-particle. It becomes clear by studying a few examples that these are the only types of combinations that create a single string of length $q$ with $p-1$ or $p+1$ particles, respectively.

For the coming analysis, it will be of use to introduce a convenient notation, letting \\\\$[\tilde C10]\ \rightarrow\ \uparrow$ and $[\tilde C01]\ \rightarrow\ \downarrow$.\\\\ In this spin-1/2 language, the fractionally charged excitation is a domain wall between the states $[\uparrow]=...\uparrow\uparrow\uparrow...$ and $[\downarrow]=...\downarrow\downarrow\downarrow...$. The resulting domain walls, e.g., $...\downarrow\downarrow{\color{red}\underline{\downarrow\uparrow}}\uparrow\uparrow...$, are characterized by one nearest-neighbor spin pair---here red-colored and underlined---having opposite polarization. This will be generalized in Section \ref{subB}.  

As an illustration of the above, consider $\nu=2/5$ and the domain wall between the states $[\tilde C01]=[\downarrow]=[01001]$ and $[\tilde C10]=[\uparrow]=[01010]$:\\\\
$[\tilde C01][\tilde C10] =[\downarrow][\uparrow]=[01001][01010]=$\\\\$...01001010010100{\color{red}\underline{10101}}00101001010...$.\\\\   
Here, the red-colored and underlined string is the only one of length $q=5$ to host three particles instead of two; hence, this is an $\frac{e}{5}$ charge.

Most famously, fractional charges $\pm\frac{e}{q}$ appear in the fractional quantum-Hall system for so-called Abelian filling fractions, including, e.g., $\nu=1/3$ and $\nu=2/5$. In these systems, which are two-dimensional in the laboratory, the excitations obey anyonic statistics \cite{leinaas}, implying that an exchange of two particles comes with a phase factor differing from $\pm1$. However, there are \emph{non-Abelian} quantum-Hall systems as well \cite{stern,mr}, with excitations believed to obey non-Abelian statistics; here, an exchange (braiding) of particles gives a new state instead of simply a phase factor. In the one-dimensional lattice studied in this paper, braiding of particles is not unambiguously defined, and neither is therefore fractional exchange statistics, see however \cite{flavin}. The states treated in this work, and the corresponding non-Abelian systems mentioned throughout the text, are related through their common exclusion statistics \cite{exclusion,eddy,micke}. Moreover, in the FQH system, these simple states are adiabatically connected to the bulk QH states, see \cite{emilPRB}.


\subsection{Non-convexity and non-Abelian states}\label{subnon}

Perhaps the most well known example of a non-Abelian system is the quantum-Hall system at half-filling in the second Landau level \cite{landau}, supporting minimal excitations of charge $\pm e/4$. On the infinitely thin torus (or, the TT limit), the so-called Moore-Read (MR) states \cite{mr}, believed to describe this system, are sixfold degenerate (instead of just twofold) and take the form of the lattice states $[10]$ (two translated copies) and $[1100]$ (four translated copies) \cite{minforsta,seidel}. The first of these is recognized as the TT state, while the other has a different symmetry. When these states are degenerate, a minimal excitation is a domain wall between the two; e.g.,\\\\ 
$[10][1100]\equiv...101010{\underline{\color{red}1011}}0011001100...$.\\\\ 
Here, the red-colored and underlined string is the only string of four sites in the state to host three particles instead of two---hence, the domain wall quasi-particle carries the charge $e/4$. By allowing for a nontrivial ground-state degeneracy, the ordinary $e/2$-charges, expected for $q=2$, split into $e/4$-charges. The increased degeneracy also implies a larger number of different ways to create an excitation at a specific lattice location, see also Section \ref{sec3}. This feature is characteristic for non-Abelian systems \cite{eddy}. 

Besides being interesting in their own right, non-Abelian systems have a proposed central role in future quantum computing \cite{nayak} and it is of general interest to frame a systematic way to explore such phases using the one-dimensional model of equation \eqref{eqn:ham}. Clearly, the increased ground-state degeneracy of, e.g., the MR states, is not supported by any convex interaction through equation \eqref{eqn:ham}. There are however other possibilities to include more states in the ground-state manifold---a statement which rests on the following: Defining the energy of the TT state to be zero, the energy of any other lattice state may be written
\begin{equation}
\label{eqn:energy}
E=\sum_{j=1}^q a_jV''_j+\sum_{j=q+1}^\infty b_jV''_j,
\end{equation}
where all $a_j$, $b_j\geq0$ and at least one $a_j>0$ \cite{comi}. Assuming $V''_j\geq0$ $\forall j$, this implies:\\

(i) If $V''_j>0$, $j\leq q$, the $q$-fold degenerate TT state is the \emph{unique} ground state of $H$.\\

(ii) If, on the other hand, $V''_j=0$ for one or several $j\leq q$, it is possible for other lattice states to have zero energy \emph{as well}, yielding an extra, nontrivial, ground-state degeneracy \cite{bosonartikeln,comii}.\\\\
Here, the formulation \emph{nontrivial degeneracy} means a degeneracy of inequivalent states, not related by center-of-mass translation, and (ii) is what one needs to exploit in order to find new exotic ground states, along with the ordinary TT states.

In the next section we present a certain class of non-convex (in the sense of fulfilling (ii)) interactions giving rise to new sets of low-energy lattice states, which agree with features of non-Abelian phases. In this picture, the old results concerning convex interactions are rediscovered as special cases of new, general findings.


\section{New ground states and fractional excitations}\label{sec3}

The procedure for calculating the values of the fractional charges, described in Section \ref{sec2}, leads to an interesting insight: if one could create a state with a single string of length $mq$ ($m$ being some positive integer) hosting $mp\pm1$ particles, while all other strings of length $mq$ host $mp$ particles, one would have achieved a fractional charge $\pm\frac{e}{mq}$. In other words, for $m>1$, one would have a splitting of the ordinary Abelian fractional charges. Moreover, it can be realized that such an excitation \emph{can} be constructed by forming domain walls between states with extended unit cells of length $mq$, built up by $m$ copies of $[\tilde C10]=[\uparrow]$ and/or $[\tilde C01]=[\downarrow]$ in any order. For $m=3$, e.g., these states have the following spin unit cells:\\\\ 
$[\uparrow\uparrow\uparrow]$, $[\uparrow\uparrow\downarrow]$, $[\uparrow\downarrow\uparrow]$, $[\downarrow\uparrow\uparrow]$, $[\uparrow\downarrow\downarrow]$, $[\downarrow\uparrow\downarrow]$, $[\downarrow\downarrow\uparrow]$ and $[\downarrow\downarrow\downarrow]$.\\\\
Note that the spin-polarized states are TT states and that they, along with other states such as $[\uparrow\downarrow\uparrow\downarrow]$, have a reducible unit cell.

In this section we explore the nature of these states and their domain wall excitations, as well as present a class of interaction parameters which actually gives rise to them.


\subsection{Degenerate ground states}\label{sub1}

Before we investigate the nature of the fractional domain walls that can be constructed using the spin states above, let us deduce the degeneracy of this new set of ground states. (For a proof that they actually are degenerate ground states for certain types of interactions, see the Appendix.) For the following analysis we introduce $S_{m,j}$ as the set of states consisting of $(m-j)$ spin-ups and $j$ spin-downs in any order. Here, $j=0,1,...,m$.
Clearly, there are $\left(\begin{array}{c}
                                             m \\
                                             j \\
\end{array}\right)$ different spin states contained in $S_{m,j}$. Summing up all the spin states then gives $\sum_{j=0}^m\left(\begin{array}{c}
                                             m \\
                                             j \\
\end{array}\right)=2^m$. However, the degeneracy we are interested in is \emph{the number of inequivalent Fock states corresponding to the spin states}, rather than the number of spin states themselves. At this stage it is appropriate to consider the fact that we could have used any of the $q$ translations of the TT state $[\tilde C10]$ as our starting point when constructing the spin subspace. Therefore, there are $q$ copies of the spin subspace, each equally valid. Concluding, the total ground-state degeneracy is 
\begin{equation} d_{q,m}=q\sum_{j=1}^m\left(\begin{array}{c}
                                             m \\
                                             j \\
\end{array}\right)=q(2^m-1),\label{dqm}\end{equation}
where $j=0$ has been excluded from the sum to take care of the fact that each TT state appear in two different spin-subspace copies---in one in the shape of $[\uparrow\uparrow...\uparrow]$ and in another in the form of $[\downarrow\downarrow...\downarrow]$. Notably, equation \eqref{dqm} implies an increased degeneracy for $m>1$ and reproduces the ordinary center-of-mass degeneracy for $m=1$.

To illustrate our results, let us look at a concrete example, namely  $\nu=2/5$, $m=2$. Here, the ground states can be divided into five copies of the spin subspace \cite{comorder};\\

$S_{2,0}^{(1)}=\{[\uparrow\uparrow]=[0101001010]\},$

$S_{2,1}^{(1)}=\{[\uparrow\downarrow]=[0101001001], [\downarrow\uparrow]=[0100101010]\},$

$S_{2,2}^{(1)}=\{[\downarrow\downarrow]=[0100101001]\},$\\

and\\

$S_{2,0}^{(2)}=\{[0100101001]\}=S_{2,2}^{(1)},$

$S_{2,1}^{(2)}=\{[0100100101]$, $[0010101001]\},$

$S_{2,2}^{(2)}=\{[0010100101]\},$\\

and\\

$S_{2,0}^{(3)}=\{[0010100101]\}=S_{2,2}^{(2)},$

$S_{2,1}^{(3)}=\{[0010010101]$, $[1010100100]\},$

$S_{2,2}^{(3)}=\{[1010010100]\},$\\

and\\

$S_{2,0}^{(4)}=\{[1010010100]\}=S_{2,2}^{(3)},$

$S_{2,1}^{(4)}=\{[1010010010]$, $[1001010100]\},$

$S_{2,2}^{(4)}=\{[1001010010]\}$,\\

and\\

$S_{2,0}^{(5)}=\{[1001010010]\}=S_{2,2}^{(4)},$

$S_{2,1}^{(5)}=\{[1001001010]$, $[0101010010]\},$

$S_{2,2}^{(5)}=\{[0101001010]\}=S_{2,0}^{(1)}$.\\\\
The inequivalent Fock unit cells are thus $[0101001010]\equiv[01010]$ (5 translations) and $[0101001001]$ (10 translations). Accordingly, the degeneracy is $d_{5,2}=5(2^2-1)=15$.


\subsection{Split fractional charges}\label{subB}

In Section \ref{subconvex} we noted that the single string of $p\pm1$ particles on $q$ sites corresponds to a single nearest-neighbor spin pair of opposite polarization. This may be directly generalized to any other value of $m$ in the following way (see the Appendix for details): an excitation with fractional charge $e^*=\pm\frac{e}{mq}$ is in the introduced spin notation characterized by  a single pair of $m$:th nearest-neighbor spins of opposite polarization, whilst all other such pairs have equal polarization. When viewed in number representation, these spin states namely have one single string of length $mq$ in which there are $mp\pm1$ particles, whereas all other strings of the same length host $mp$ particles. 

It is quickly realized that this type of spin domain wall is achieved by combining spin states related through the flipping of one spin in the spin unit cell, i.e., by combining states in $S_{m,n}$ with certain states in $S_{m,n\pm1}$. The domain wall $[S_{m,n}][S_{m,n\pm1}]$ corresponds to a fractional charge $e^*=\mp\frac{e}{mq}$. Note, though, that all states in $S_{m,n\pm1}$ will not do---the rule is to connect states related by flipping one spin. For example, for $m=5$, the state $[\uparrow\downarrow\downarrow]$ in $S_{3,2}$ may be combined with $[\uparrow\uparrow\downarrow]$ in $S_{3,1}$---though not with $[\downarrow\uparrow\uparrow]$ in $S_{3,1}$, since these two are not related by the flipping of one single spin. Notably, each state within $S_{m,n}$ may be combined with $n$ different states in $S_{m,n-1}$ and $m-n$ different states in $S_{m,n+1}$.  

\begin{figure}[h!t]
\begin{tikzpicture}
			\node (S01) {$\left[S_{3,0}^{(1)}\right]$}; 
			\def\lastx{0}\def\lasty{1}
			\foreach \x/\y [remember=\x as \lastx, remember=\y as \lasty] in {1/1,2/1,0/2,1/2,2/2,0/3}{%
				\node (arrow\lastx\lasty) [below=0pt of S\lastx\lasty] {$q_p\uparrow\downarrow q_h$};
				\node (S\x\y) [below=0pt of arrow\lastx\lasty] {$\left[S_{3,\x}^{(\y)}\right]$};
				}
			\node (arrow\lastx\lasty) [below=0pt of S\lastx\lasty] {$q_p\uparrow\downarrow q_h$};
			\node (dots) [below=10pt of S\lastx\lasty] {$\vdots$};
			\node (arrowbelow) [below=0pt of dots] {$q_p\uparrow\downarrow q_h$};
			\node (S2q) [below=0pt of arrowbelow] {$\left[S_{3,2}^{(q)}\right]$};
			\node (top) [above=5pt of S01] {};
			\node (topleft) [left=20pt of top] {};
			\node (bottom) [below=5pt of S2q] {};
			\node (bottomleft) [left=20pt of bottom] {};
			\draw (S2q)-- (bottom.center)-- node (bottomarrow) {\midarrow} (bottomleft.center)--(topleft.center)-- node (toparrow) {\midarrow} (top.center)--(S01) ; 
			\node [above=0pt of toparrow] {$q_p$};
			\node [below=0pt of bottomarrow] {$q_h$};
			\matrix (m)
	 		 [right=0pt of S21, matrix of nodes,%
    		 		nodes in empty cells,
    		 		nodes={outer sep=0pt,circle,minimum size=10pt},
    		 		column sep={1.35cm,between origins},
  		 		 row sep={1.35cm,between origins}]
			{
				&&&&\\
				$n=0$&$1$&$2$&$3$&$\cdots$\\
				&&&&$\cdots$\\
				&&&&\\
				&&&&$\cdots$\\
				&&&&\\
				&&&&$\cdots$\\
				&&&&\\
				&&&&$\cdots$\\
			};
			\draw[->,thick] (m-6-1.center) -- (m-5-2.center);
				\draw[->,thick] (m-5-2.center) -- (m-4-3.center);
					\draw[->,thick,thick] (m-4-3.center) -- (m-3-4.center);
					\draw[->,thick] (m-4-3.center) -- (m-5-4.center);
				\draw[->,thick] (m-5-2.center) -- (m-6-3.center);	
					\draw[->,thick] (m-6-3.center) -- (m-5-4.center);
					\draw[->,thick] (m-6-3.center) -- (m-7-4.center);
			\draw[->,thick] (m-6-1.center) -- (m-7-2.center);	
				\draw[->,thick] (m-7-2.center) -- (m-6-3.center);
				\draw[->,thick] (m-7-2.center) -- (m-8-3.center);
					\draw[->,thick] (m-8-3.center) -- (m-7-4.center);
					\draw[->,thick] (m-8-3.center) -- (m-9-4.center);									
	
\end{tikzpicture}
\caption{Diagram illustrating the formation of fractional charges as domain walls between degenerate ground states (the levels in the left column) for $m=3$. Going from left to right, a domain wall is created in each step; the integer $n$ denotes the number of excitations. In this particular example, the grid illustrates the various ways to create domain walls starting from a state in $S_{3,0}^{(2)}$. Each ``clockwise" step creates a quasi-hole ($e^*=-\frac{e}{mq}$), whereas every ``anti-clockwise" step creates a quasi-particle ($e^*=\frac{e}{mq}$), as indicated by the arrows between different levels.}
\label{bratteli}
\end{figure}
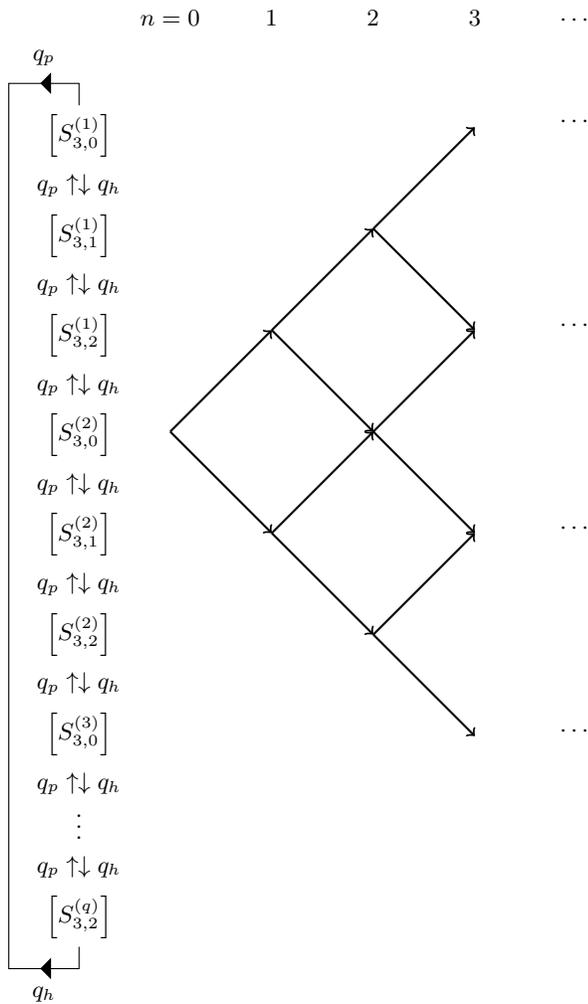

Let us illustrate the domain wall construction by making use of the example treated in Section \ref{sub1}. For $\nu=2/5$, $m=2$, we can create an excitation by combining, e.g., $[\downarrow\downarrow]=[0100101001]$ and $[\downarrow\uparrow]=[0100101010]$:\\\\
$[\downarrow\downarrow][\downarrow\uparrow]=...\downarrow\downarrow\downarrow\downarrow\downarrow{\color{red}\underline{\downarrow}}\downarrow{\color{red}\underline{\uparrow}}\downarrow\uparrow\downarrow\uparrow...=$\\\\
$...0100101001010010100{\color{red}\underline{1010010101}}00100101010...\,$.\\\\
Here, we have colored and underlined the two opposite spins and the fractionally charged ($e^*=e/10$) string.

To construct sequences of more than $q-1$ consecutive quasi-particles or quasi-holes will require the inclusion of states from different spin subspaces. Note that the TT state $\big{[}S_{m,m}^{(i)}\big{]}$ in one subspace is equal to $\big{[}S_{m,0}^{(i+1)}\big{]}$ in some other subspace, see example in Section \ref{sub1}, which means that $\big{[}S_{m,0}\big{]}^{(i)}\big{[}S_{m,m-1}^{(i+1)}\big{]}=\big{[}S_{m,m}^{(i+1)}\big{]}\big{[}S_{m,m-1}^{(i+1)}\big{]}$ will make a quasi-particle, and so on.
An example of a state with six quasi-holes, $m=3$, is\\\\
$\big{[}S_{3,0}^{(i)}\big{]}\big{[}S_{3,1}^{(i)}\big{]}\big{[}S_{3,2}^{(i)}\big{]}\big{[}S_{3,0}^{(i+1)}\big{]}\big{[}S_{3,1}^{(i+1)}\big{]}\big{[}S_{3,2}^{(i+1)}\big{]}\big{[}S_{3,0}^{(i+2)}\big{]}$.\\\\

The creation of domain wall excitations is advantageously illustrated by a so-called Bratteli diagram \cite{bratteli}, see Figure \ref{bratteli}, where the construction of a domain wall is represented by drawing a line from one state to an adjacent one in the diagram. Here, a step in the downward direction corresponds to a quasi-hole, and a step in the upward direction corresponds to a quasi-particle. Note that there are many inequivalent ways to create $n$ quasi-particles and/or quasi-holes. Each level of states, denoted $[S_{m,j}]$ in the column to the left, represents $\left(\begin{array}{c}
                                             m \\
                                             j \\
\end{array}\right)$ different Fock states. The many inequivalent ways to create a set of domain walls are intimately connected to the concept of non-Abelian statistics, see, e.g., Ref. \cite{eddy}.


\subsection{Interaction tuning and experimental implications}\label{subexp}

A question of obvious interest is whether it is possible in principle to specify an interaction, defined by the values of $V''_j$, that actually would produce the low-energy physics described above. This is indeed the case, as is shown in some detail in the Appendix. There, we make an appropriate departure from convexity by imposing $V''_{kq}=0$, $k$ integer - with the exception $k=m$, $m$ being the positive integer introduced above; $V''_{mq}$ should be positive but small. We find that, under these circumstances, the spin states are degenerate ground states and the minimal excitations, carrying fractional charges of size $e^*=\pm\frac{e}{mq}$, are constructed as domain walls between the degenerate vacua, as those exemplified in Section \ref{subB}. As an example, let us return to the Moore-Read system ($q=2$) with its fractional charges $e^*=\pm\frac{e}{4}$. Using our gained knowledge, we should impose $m=2$ in this case, implying $V''_{2k}=0$, $k\neq 2$, and $V''_{4}>0$. In other words, $V''_j$, $j$ even, should be shrunk to zero---except for $V''_4$ which should be small but finite. For this type of interaction, the six MR states are degenerate ground states of the Hamiltonian in \eqref{eqn:ham} \cite{bosonartikeln} and the low-energy excitations are domain walls of fractional charge $e^*=\pm\frac{e}{4}$ between them.

The prospects of realizing or mimicking these restricted interaction parameters experimentally might first be considered in the context of ultracold gases in optical lattices. Although today's experimental techniques allow for tuning the on-site interaction (and hence $V''_1$) through Feshbach resonances \cite{feshbach}, methods which enable tuning of every $V''_j$ individually is out of reach in current schemes. An alternative, more experimentally relevant outlook for achieving the fractional domain walls is offered by directly adjusting the chemical potential for each lattice site separately by means of single-site addressing \cite{bakr,sherson}. Using such ideas, the number of particles on each individual site may be controlled to create a specific domain wall of desired charge, in effect mimicking the non-convex interaction which would result in the same kind of lattice state. Reflecting on the possibility to detect non-Abelian statistics in general, we observe the recent scheme suggesting that braiding of quasi-particles, which inevitably requires more than one dimension, might be possible through the construction of a T-junction configuration \cite{alicea}. An optical lattice naturally contains sites in the transverse directions that can be utilized for constructing the T-junctions. Using this and changing the on-site chemical potential in the optical lattice, one can envision similar braiding in the present system.   

Even in the absence of a fine-tuned interaction, the nontrivially degenerate spin states might well be of relevance from an experimental point of view. It is interesting to note that some of these states correspond to quantum-Hall states---Abelian and non-Abelian---which demonstrably exist in nature and do not require interaction fine-tuning since they are robust against perturbations. This is suggestive for that other states emerging from this analysis might be related to QH states not yet investigated. Also, some states emerging from our analysis correspond to QH states which are less investigated experimentally, but the more studied theoretically, such as the so-called Read-Rezayi states \cite{rr}, which are non-Abelian. It would furthermore be interesting to compare the states analyzed in Refs. \cite{wen,bernevig} with the more generic ones found using our approach.  

As a final remark, let us mention that entering a tunneling term into equation \eqref{eqn:ham} breaks the degeneracy of the inequivalent ground states. Because tunneling can be made very small, it might be possible to compensate for this by shifting the requirements on $V''_j$ slightly, achieving a pairwise ground-state degeneracy, as argued in \cite{jonas}. However, it will not be possible to achieve an exact $q(2^m-1)$-fold degeneracy. One suggestion might be that because of the many inequivalent ways to create the split fractionally charged domain walls, those could nevertheless be seen in experiments due to the minimization of the free energy in finite-temperature systems.


\section{Summary}\label{sec5}

We have investigated a 1D lattice model, relevant to diverse areas like the fractional quantum Hall effect, TCNQ salts and cold-atom systems. With the objective to find sets of nontrivially degenerate ground states, as well as exotic low-energy excitations, we consider general filling fractions $\nu=p/q$ and a certain non-convex set of interaction parameters $V''_j$, defined by the positive integer $m$. For this interaction, we find a $q(2^m-1)$-fold degenerate ground-state manifold and an excitation structure of fractional charges $e^*=\pm\frac{e}{mq}$, formed as domain walls between qualitatively different ground states. In other words, fractional charges of more general values than just $e^*=\pm\frac{e}{q}$ (here appearing as the special case $m=1$) may emerge. The many inequivalent ways to create the domain walls are intimately connected to the concept of non-Abelian statistics, which has important applications in future quantum-computational schemes. Though the formal analysis rests on a fine-tuned interaction, we suggest that our results have more general bearing due to the connection to known non-Abelian quantum-Hall states which do not require fine-tuning. Single-site addressing in optical lattices might be suitable for realizing the fractionally charged domain walls in an experimental setting.  


\begin{acknowledgements}
The author much appreciates valuable discussions and collaboration with Emil J. Bergholtz, Mikael Fremling, Anders Karlhede, and Jonas Larson. Acknowledgments to Thomas Kvorning for contributing with the making of Figure \ref{bratteli}.
\end{acknowledgements}


\section*{APPENDIX}

Here we present some of the details behind the results presented in this paper. The formal analysis is conducted for rational filling fractions $\nu=p/q$ with the exceptions of $\nu=n$, $\nu=\frac{1}{q}+n$, and $\nu=\frac{q-1}{q}+n$, $n=0, 1, 2,...$ \cite{com2}. The excepted fillings will be treated qualitatively later on. 


\subsection{Mapping of low-energy sector onto spin-1/2 chains}

The aim is to search for ground states and minimal excitations of the Hamiltonian in \eqref{eqn:ham} for certain choices of interaction. We first assume that the cost $V_0$ of a pair of particles occupying the same site is so large that this is avoided. In practice, this means that, for $\nu<1$, the particles behave as fermions, while for $\nu>1$ we may treat them as fermions at filling $\nu<1$ on top of a uniform, inert background of integer filling. Below, then, we have subtracted any integer-filling background and treat every system like fermions at $\nu<1$. Under this circumstance, the Hamiltonian reduces to
\begin{equation}
\hat H_f=\sum_{i,j=1}^\infty V_{j}\hat n_i\hat n_{i+j}
\label{hf}
\end{equation}
and we treat $1<p<q-1$. For these filling fractions, the following holds \cite{micke}:\\ 

(I) The unit cell of the TT state may be written $[\tilde C10]$, where $\tilde C$ is a binary string, mirror symmetric around its center, see also Section \ref{subconvex}. For $\nu=2/5$, $\tilde C=010$, so that $[\tilde C10]=[01010]$.\\

(II) Within each cell $[\tilde C10]$ there are two particles which may be moved to create a translation of the original TT unit cell. The rightmost particle may be moved one step to the right; $[\tilde C10]\rightarrow[\tilde C01]$, and there is also one particle inside $\tilde C$ that may be moved one step to the left. Using our previous example, the first operation is $[01010]\rightarrow[01001]$, and the latter operation translates into $[01010]\rightarrow[10010]$.\\ 

(III) The cost of performing one of the operations of (II) in a cell in the TT state is $V''_q+V''_{2q}+V''_{3q}+...$. On the other hand, the cost of any other operation is of the order $V''_{j<q}$ \cite{com3}.\\  

From (III) it follows that whenever $V''_{j<q}>V''_{j\geq q}$, the low-energy sector of the system consists of states containing sequences of cells of length $q$, in which each of the two ``movable" particles occupies either of its two possible sites. This low-energy Hilbert space will be denoted $\mathcal H'$ and because of the $q$-fold center-of-mass degeneracy of the TT state, there are $q$ equally valid translated copies of $\mathcal H'$. 

Considering $V''_{j<q}>V''_{j\geq q}$, we will now denote each cell in a state in $\mathcal H'$, by two spins, $s'_i$ and $s_i$, which may take the values $\pm\frac{1}{2}$ (``up" and ``down"). The spins are given by the positions of the movable particles in cell $i$, according to $10\ \rightarrow\ \ \uparrow$, and $01\ \rightarrow\ \ \downarrow$. For example, the low-energy subspace of $\nu =2/5$ consists of states constructed out of the cells $[01010]\equiv(\downarrow\uparrow)$, $[01001]\equiv(\downarrow\downarrow)$, $[10010]\equiv(\uparrow\uparrow)$ and $[10001]\equiv(\uparrow\downarrow)$. Note that this is not yet the same notation as is being used in the main text, where each cell is characterized by one single spin---see what follows below. The combinations $(s'_i, s_i)=(\downarrow\uparrow)=[\tilde C10]$, $(\downarrow\downarrow)=[\tilde C01]$ and $(\uparrow\uparrow)$ are all translations of the TT unit cell, while  $(\uparrow\downarrow)$ is not. Due to this, the latter type of cell will not appear in the lowest energy states.  

Now, our goal is to find the low-energy states of the Hamiltonian in equation \eqref{hf}, exploiting the spin-1/2 mapping of the low-energy subspace $\mathcal H'$. We thus need to translate the number operators $\hat n_i$, for acting within $\mathcal H'$, into spin-operators. This is a straightforward procedure which leads to the following expression:
\begin{equation}
\begin{array}{lll}
\hat H'=&\sum_{i,k=1}^{\infty}& \displaystyle{\Big{\{}-V''_{kq}(s'_i s'_{i+k}+s_i s_{i+k})}\\\\ 
& & -\displaystyle{V''_{kq-l_p}(s'_is_{i+k-1}-\frac{1}{2}s'_i+\frac{1}{2}s_i)}\\\\
& & -\displaystyle{V''_{kq-l_h}(s_is'_{i+k}-\frac{1}{2}s'_i+\frac{1}{2}s_i)\Big{\}},}
\end{array}
\label{hprim}
\end{equation}
where $l_p$ and $l_h$ are related to the filling fraction, and $l_p+l_h=q$ \cite{comlp}.
This expression becomes somewhat simplified once we impose a special requirement on the interaction: $V''_{kq}=0$, $k\neq m$, $V''_{mq}>0$, where $m$ is some positive integer. Remember that, according to (ii) in Section \ref{subnon}, imposing $V''_q=0$ opens for the creation of nontrivially degenerate ground states and note that this holds whenever $m>1$. Equation \eqref{hprim} turns into 
\begin{equation}
\begin{array}{lll}
\hat H''=&\sum_{i=1}^\infty&\displaystyle{\Big{\{}-V''_{mq}(s'_i s'_{i+m}+s_i s_{i+m})}\\\\ 
& & \displaystyle{+\sum_{k=1}^{\infty} -V''_{kq-l_p}(s'_is_{i+k-1}-\frac{1}{2}s'_i+\frac{1}{2}s_i)}\\\\
& & \ \ \ \ \ \ \ \ \ \ \displaystyle{-V''_{kq-l_h}(s_is'_{i+k}-\frac{1}{2}s'_i+\frac{1}{2}s_i)\Big{\}}}.
\end{array}\label{hbis}
\end{equation}


\subsection{Ground states and minimal excitations}

The ground states and minimal excitations of $\hat H''$ may now readily be obtained. First note that the term containing $V''_{mq}$ is minimized by all spin states with periodicity $m$, i.e., states for which $s'_i=s'_{i+m}$ and $s_i=s_{i+m}$ $\forall i$. The TT state obviously falls within this category and when $m=1$, the TT state (being the only state with $s'_i=s'_{i+1}$, $s_i=s_{i+1}$ $\forall i$) is the unique ground state. (This should not come as a surprise, since $m=1$ implies that relation (i) in Section \ref{subnon} is fulfilled.) More interesting is $m>1$, for which there are other states than the TT state with (spin) periodicity $m$. 

Next, considering the other terms in \eqref{hbis}, one finds that there are spin states with periodicity $m$ which minimize also the terms proportional to $V''_{kq-l_p}$ and $V''_{kq-l_h}$ $\forall\ i,k$. Investigating the possible combinations of pairs of $s'_i$ and $s_{i+k-1}$, and $s_i$ and $s'_{i+k}$, respectively, one finds that these terms are minimized by the spin states where either $s'_i=\downarrow$ everywhere, \emph{or else}, $s_i=\uparrow$ everywhere (if one of these requirements are fulfilled, the other spins can take any value). These two alternatives are equivalent in the sense that the states fulfilling $s'_i=\downarrow$ $\forall i$ in one copy of $\mathcal H'$, appear as the states fulfilling $s_i=\uparrow$ $\forall i$ in another translated copy of $\mathcal H'$. In other words, the two types of states are trivially connected through the center-of-mass translation, which means that we, in practice, need only to consider \emph{one} spin variable per unit cell---say, e.g., $s_i$---without loss of generality. Our ground states may hence be described solely by the spins $s_i$. Please note the potentially confusing shift in notation; from now on, $[\tilde C 10]\equiv\uparrow$ and $[\tilde C 01]\equiv\downarrow$. This is the notation introduced in and used throughout the main text. 

To conclude: In the introduced notation, the ground-state manifold consists of all states constructed by $[\tilde C 10]\equiv\uparrow$ and $[\tilde C 01]\equiv\downarrow$, having periodicity $m$---i.e. all states with a 1D spin-1/2 unit cell of length $m$. 

Let us turn our attention to the excitation spectrum of $\hat H''$. Assuming that $V''_{mq}$ is small compared to other $V''_j$'s in equation \eqref{hbis}, the smallest excitation possible corresponds to a state where all pairs of spins $s_i$, $s_{i+m}$, \emph{except one} such pair, have the same sign. Such states are in turn created by constructing domain walls between states with $j$ (see section \ref{sub1}) differing by 1. The rest of the analysis can now be followed in the main text. 


\subsection{Generalization to other filling fractions}

The cases $\nu=\frac{1}{q}+n$ and $\nu=\frac{q-1}{q}+n$ differ slightly from the ones treated in the previous parts of the Appendix. When $p=1$ (integer background fillings being ignored, as before), there is only \emph{one} particle per TT unit cell, able to move one step to the left \emph{or} one step to the right at the cost of $\sim V''_q$. Analogously, when $p=q-1$ there are two movable particles in the unit cell but they share one of their available sites, in effect enabling the single empty site to move one step to the left or to the right. These fillings may not immediately be incorporated in the spin-$1/2$ mapping, since moving in those cases involves three sites instead of four. However, a few examples suggest that things work analogously when it comes to picking ground states and excited states for the same choices of $V''_j$ as before. To be specific: the lowest energy states will be built up by the cells $[\tilde C 10]=\uparrow$ and  $[\tilde C 01]=\downarrow$, where $\tilde C$ is a string of $q-2$ empty (filled) sites for $p=1$ ($p=q-1$). Filling $\nu=1/2$ is a special case, where $\tilde C$ has zero extension. 

As mentioned before, integer fillings $\nu=n$ are more special yet in the sense that, no matter the strength of the on-site interaction $V_0$, one needs to consider states containing sites hosting more than one particle---simply because, in this case, there is only one state without this feature. Inevitably, excitations will involve multiple occupied sites \cite{com5}. For convex interactions, the TT state ($n$ particles on every site) is the unique ground state and a minimal excitation is created by moving one particle one step in the lattice---e.g., $...22222222...\rightarrow...22231222...$. No charge fractionalization occurs here; to achieve a splitting of charges, what is needed, again, is a departure from convexity. More particular, since $q=1$, one will have to require $V''_1=V_0+V_2-2V_1=0$. Qualitatively, this means that the energy cost for each on-site pair of particles (or, the cost of having a pair of particles two lattice sites apart) must be diminished. By making suitable choices for the interaction parameters, fractional charges $e^*=\pm\frac{e}{2}$ may be created through an increased ground-state degeneracy. For details on this, we refer to \cite{jonas}.



\begin{thebibliography}{999}

\bibitem{hubbard}J. Hubbard, Phys. Rev. B {\bf 817}, 494 (1978).
\bibitem{bloch}I. Bloch, J. Dalibard, and W. Zwerger, Rev. Mod. Phys. {\bf 80}, 885 (2008).
\bibitem{lewenstein}M. Lewenstein, A. Sanpera, V. Ahufinger, B. Damski, A. Sen(de), and U. Sen, Adv. Phys. {\bf 56}, 243 (2007).
\bibitem{com1}In other words, assuming that the kinetic energy is negligible compared to the inter-particle interaction---hence the formulation "strong non-convex interactions" in the title.
\bibitem{tsui}D. C. Tsui, H. L. Stormer, and A. C. Gossard, Phys. Rev. Lett. {\bf 48}, 1559 (1982). 
\bibitem{emilPRB}E. J. Bergholtz, and A. Karlhede, Phys. Rev. B {\bf 77}, 155308 (2008).
\bibitem{susanne}S. Viefers, J. Phys.: Condens. Matter {\bf 20}, 123202 (2008).
\bibitem{emilPRL}E. J. Bergholtz, T. H. Hansson, M. Hermanns, and A. Karlhede, Phys. Rev. Lett. {\bf 99}, 256803 (2007).
\bibitem{pokrovsky}V. L. Pokrovsky, and G. V. Uimin, J. Phys. C: Solid State Phys. {\bf 11}, 3535 (1978).
\bibitem{burnell}F. J. Burnell, M. M. Parish, N. R. Cooper, and S. L. Sondhi, Phys. Rev. B {\bf 80}, 174519 (2009).
\bibitem{bosonartikeln}E. Wikberg, E. J. Bergholtz, and A. Karlhede, J. Stat. Mech., P07038 (2009).
\bibitem{jonas}E. Wikberg, J. Larson, E. J. Bergholtz, and A. Karlhede, Phys. Rev. A {\bf 85}, 033607 (2012).
\bibitem{bauer}M. Bauer, and M. M. Parish, Phys. Rev. Lett. {\bf 108}, 255302 (2012).
\bibitem{su}W. P. Su, and J. R. Schrieffer, Phys. Rev. Lett. {\bf 46}, 738 (1981).
\bibitem{stern}A. Stern, Nature {\bf 464}, 187 (2010).
\bibitem{nayak}C. Nayak, S. H. Simon, A. Stern, M. Freedman, and S. Das Sarma, Rev. Mod. Phys. {\bf 80} 1083 (2008).
\bibitem{mr}G. Moore, and N. Read, Nucl. Phys. B {\bf 360}, 362 (1991).
\bibitem{tt}R. Tao, and D. J. Thouless, Phys. Rev. B {\bf 28}, 1142 (1983).
\bibitem{eddy}E. Ardonne, E. J. Bergholtz, J. Kailasvuori, and E. Wikberg, J. Stat. Mech., P04016 (2008).
\bibitem{comCtilde}This holds for $0<\nu<1$. In the Appendix we comment on how the analysis may be generalized to other fillings as well.
\bibitem{micke}M. Kardell, and A. Karlhede, J. Stat. Mech., P02037 (2011).
\bibitem{leinaas}J. M. Leinaas, and J. Myrheim, Il Nuovo Cimento B {\bf 37} (1), 1-23 (1977).
\bibitem{flavin}J. Flavin, and A. Seidel, Phys. Rev. X {\bf 1}, 021015 (2011).
\bibitem{exclusion}F. D. M. Haldane, Phys. Rev. Lett. {\bf 67}, 937 (1991).
\bibitem{landau}L. D. Landau, and E. M. Lifshitz, \emph{Quantum Mechanics}, Pergamon Press, New York (1994).
\bibitem{minforsta}E. J. Bergholtz, J. Kailasvuori, E. Wikberg, T. H. Hansson, and A. Karlhede, Phys. Rev. B {\bf 74} 081308(R) (2006).
\bibitem{seidel}A. Seidel, and D.-H. Lee, Phys. Rev. Lett. {\bf 97}, 056804 (2006).
\bibitem{comi}See, e.g., Ref. \cite{emilPRB}.
\bibitem{comii}If some $V''_j<0$, the TT state is not necessarily one of the ground states anymore. This case is not treated in this paper.
\bibitem{comorder}The ordering of the subspaces is chosen so that $\big{[}S_{m,m}^{(i)}\big{]}=\big{[}S_{m,0}^{(i+1)}\big{]}$. 
\bibitem{bratteli}O. Bratteli, Trans. Am. Math. Soc. {\bf 171}, 195 (1972).
\bibitem{feshbach}D. B. Dinkerscheid, U. Al Khawaja, D. van Oosten, and H. T. Stoof, Phys. Rev. A {\bf 71}, 043604 (2005).
\bibitem{bakr}W. S. Bakr, J. I. Gillen, A. Peng, S. Fšlling, and M. Greiner, Nature {\bf 462}, 74 (2009)
\bibitem{sherson}J. F. Sherson, C. Weitenberg, M. Endres, M. Cheneau, I. Bloch, and S. Kuhr, Nature {\bf 467}, 68 (2010).
\bibitem{alicea}J. Alicea, Y. Oreg, G. Refael, F. von Oppen, and M. P. A. Fisher, Nature Physics {\bf 7}, 412 (2011).
\bibitem{rr}N. Read, and E. H. Rezayi, Phys. Rev. B {\bf 59}, 8084 (1999).
\bibitem{wen}X.-G. Wen, and Z. Wang, arxiv:1203.3268v1 [cond-mat.str-el] (2012).
\bibitem{bernevig}B. A. Bernevig, and F. D. M. Haldane, Phys. Rev. Lett. {\bf 101}, 246806 (2008).
\bibitem{com2}The exceptions cover all fillings which cannot be incorporated in the spin model. For $\nu=\frac{1}{q}+n$, there is only one movable particle in each TT unit cell. For $\nu=\frac{q-1}{q}+n$, there are two movable particles, but they share one of the available sites and so they may not move independently of each other. Integer fillings are special in the sense that any excitation unavoidably involves increasing the number of on-site pairs.
\bibitem{com3}Moving \emph{both particles in the same cell} also costs an energy of the order $V''_{j<q}$. This is reflected by the fact that this operation appears together with the coefficients $V''_{q-l_p}$ and  $V''_{q-l_h}$ in \eqref{hprim} and \eqref{hbis}.
\bibitem{comlp}More precisely, $l_p$ and $l_h$ are the lengths of the quasi-particle and quasi-hole building blocks used to construct the TT unit cell in the hierarchy picture of the quantum-Hall system, see Ref. \cite{micke}. 
\bibitem{com5}For fermions, integer fillings are particularly uninteresting, since there is only one allowed state in the lattice system in this case---if not, of course, one studies systems where each site may host particles of different quantum numbers related to other properties, like spin.

\end{thebibliography}
\end{document}